# Effect of polarization forces on atom deposition on a non-spherical nanoparticle. Monte Carlo simulations

V. Nemchinsky[*], A. Khrabry[**]

[*] Keiser University Fort Lauderdale FL 33309 USA
[**] Princeton Plasma Physics Lab. NJ USA

**Abstract**

Trajectories of a polarizable species (atoms or molecules) in the vicinity of a negatively charged nanoparticle (at a floating potential) are considered. The atoms are pulled into regions of strong electric field by polarization forces. The polarization increases the deposition rate of the atoms and molecules at the nanoparticle. The effect of non-spherical shape of the nanoparticle is investigated by the Monte Carlo method. The shape of the non-spherical nanoparticle is approximated by an ellipsoid. Total deposition rate and its flux density distribution along the nanoparticle surface are calculated. It is shown that the flux density is not uniform along the surface. It is maximal at the nanoparticle tips.

## 1. Introduction

Low-temperature plasma is a commonly used medium to synthesize carbon and other nanoparticles (NP). In that plasma, a NP, once being created, is a subject of continuous fluxes of neutral and charged species depositing on it. First of all, highly mobile electrons charge it to a floating potential $V_{float}$

$$V_{float} = \frac{T_e}{2}\ln\left(\frac{M}{m}\right),$$

so that fluxes of ions and electrons balance each other. Here, $T_e$ is the electron temperature and M and m are the masses of ion and electron, respectfully. After that, the electric potential of a NP doesn't change much during its further growth. In most cases, the plasma degree of ionization is low, so that vast majority of species depositing at the NP are neutral atoms or molecules [1-5].

Molecules and atoms (hereafter just atoms) polarize as they move in the electric field of a charged nano-particle. Being polarized, the molecule is pulled into the strongest field locations: $F = \alpha\, \mathrm{grad}(E^2)$, where $\alpha$ is polarizability of the particle [6]. As a result, the rate of accumulation increases in comparison to the case of non-polarized particles.

Electric field distribution close to the charged body depends on its shape. The case of a spherical NP has been recently considered by M. Shneider [7, 8]. One might expect that the effect could be different for a long NP for two reasons. First, in the case of a long NP, the electric field extends for a longer distance (decreasing logarithmically for long molecules), which probably increases the effect. On another hand, an electric field has lower gradients, a factor that works in the opposite direction. It is difficult to predict what tendency will prevail. It is specifically interesting to consider the case of not very long NP because it could shed some light on the early stages of NP growth. In this note we consider how a polarizing force affects growth rate of non-spherical NP.

Our goal is to compute the deposition rate of the atoms. While for a spherical NP the deposition rate depends on a single parameter (the impact parameter p), in the case of a non-spherical NP, the deposition rate depends also on the direction of an incident atom. In this case the Monte Carlo (MC) method is the most appropriate.

## 2. Ellipsoid-like nanoparticle

This note treats the case of a small NP, much smaller than the atom or molecule free path $\lambda_a$, so that motion of the atoms/molecules can be considered neglecting their collisions. Also, it is supposed that the degree of plasma ionization is low so that the main depositing species at the NP are neutral atoms. In this situation, the only force acting on an atom is the polarization force. One can assume that the drain of these atoms doesn't disturb the plasma at a distance of about a few lengths of the NP. This means that one may simulate their motion starting from some distance, which on one hand is less than the mean free path of atoms and, on the other hand, substantially larger than the NP dimensions. It is assumed that the size of the NP is substantially less than the Debye length $\lambda_D$ so that the electric field of the NP is not screened. In other words, we consider the following hierarchy of the characteristic lengths: NP size $\ll \lambda_D \ll \lambda_a$.

Distribution of the electric field at the proximity of a NP could be approximated by the field of an ellipsoid. The solution of the corresponding electrostatic problem can be found in [6]. The distribution of the driving force of atom/molecule motion in proximity to the NP (the gradient of square of the electric field), is shown in Fig.1. In this picture, the z axis is directed along the major axis of the ellipsoid; z =0 corresponds to its center. Radial force, as one can see, tends to push an atom to the NP center (z=0) where the electric field is stronger. Axial force is directed toward to the tip of the NP.

When computing an atom's trajectory, 3D in nature, we used the fact that in view of axial symmetry along the major ellipse axis (z-axis), the projection of the angular momentum on the z-axis is conserved: $L_z = mr^2 \dot{\phi}$ = const. We wrote the atom motion equation using cylindrical coordinates. The radial component of the atom acceleration $\ddot{r} - r\dot{\phi}^2$ was, therefore, reduced to

$$\ddot{r} - \frac{L_z}{m^2 r^3},$$

axial component is obviously $\ddot{z}$.

The following non-dimensional parameters were introduced. The NP potential was measured in the floating potential $V_{float}$: non-dimensional potential of the NP was set equal to unity. NP has the form of an ellipsoid with major and minor semi-axes A and B, respectively. Electric field was measured in $V_{float}/eB$. Unit of time was $B/v_0$, where $v_0$ is the atom's velocity away from the NP. $\mu$ is dimensionless z-component of the angular momentum $L_z$ (dimensional momentum is $mv_0\,p$, and p is the impact parameter, see Fig.3).

In the non-dimensional form, the equations of motion are as follows:

$$\frac{d^2 r}{dt^2} = \beta \frac{d}{dr}(E_r^2 + E_z^2) + \frac{\mu^2}{r^3} \tag{1}$$

$$\frac{d^2 z}{dt^2} = \beta \frac{d}{dz}(E_r^2 + E_z^2) \tag{2}$$

The amount of the considered effect is determined by the parameter

$$\beta = \frac{\alpha}{mv_0^2}\left(\frac{V_{float}}{eB}\right)^2, \tag{3}$$

which is the polarization potential to the thermal energy ratio [7, 8]. Note that equations (1,2) describe the atomic trajectory without involving the Orbital Motion Limited (OML) approximation [9].

In this paper, we study the effect of polarization forces on the atomic flux to the nanoparticle surface. We compare the ratio of two fluxes: G, the flux of polarizable atoms to the NP and $G_0$, the flux of the atoms with polarization forces switched off. Also, we are interested how the flux G is distributed along the NP surface.

## 3. Parameter β estimation

Let us estimate β value for the following set of parameters. Consider carbon atoms at 300K temperature ($kT_a$ = 4.1 $10^{-21}$ J). Carbon atoms polarizability α = 2 $10^{-40}$ SI units [10]. The difference in masses of electrons and carbon ions yields for the floating potential $V_{float}$ = 5.0 $kT_e$. For electron temperature 2eV, $V_{float}$ =10eV =1.6 $10^{-18}$ J. Let the minor semi-axis of the NP be B = 2 $10^{-9}$m. For this set of parameters we obtain β=0.61. For a higher temperature of atoms, say, $T_a$ = 1000K, β is lower (0.18). It is also lower for larger NP. It is larger for carbon molecules $C_2$ and $C_3$. The latter is important since carbon vaporizes mostly in the form of molecules [11].

## 4. Examples of trajectories

Before proceeding to results of the MC simulation, we would like to present a few calculated trajectories. Atoms' trajectories were computed by solving equations (1, 2) with the Runge-Kutta method. Starting points of the trajectories were at a distance a few times larger than the major semi-axis A of the ellipsoid. The geometry used to simulate atom trajectories is shown in Fig.2.

Projection of the atom trajectory at z=0 plane are schematically displayed in Fig.3, where the impact parameter p is shown. A couple of these trajectories are shown in Fig.4. In this figure, the trajectories start at $z_0$=A/2 (half of a major semi-axis above the mid-plane) with zero initial axial speed (θ=0) but at different velocity deviations from the radial direction (different ϕ). The latter results in different impact parameters p. It can be seen that at small impact parameters the trajectories reach the NP, while at larger ϕ (larger impact parameter), the trajectory misses the NP. Simulations with different angles allowed us to find the critical angle (critical impact parameter) that separates trajectories that hit the ellipse from those that turn away from the NP without reaching it. Dependence of this so defined critical parameter on the initial speed of an atom (far from NP) is shown in Fig. 5. One can see rather strong dependence on the initial speed, however, this effect could be expected in view of the structure of the parameter β.

## 5. Monte Carlo procedure

The atoms' trajectories were monitored as the atoms move inside of a sphere of a radius several times larger than major semi-axis of the elliptic NP. The trajectories started at some randomly picked point uniformly distributed over the surface of this large sphere[1]. Also, randomly picked were the cosine of the angle θ and the angle ϕ of the start atom velocity, see Fig. 2. Most simulations were made in a single velocity approximation: it was assumed that far from the NP all the atoms have the same absolute value of velocity. In a few runs, however, maxwellian distribution of the velocities were assumed, see below.

The trajectories were traced until the atom reaches the NP surface or leaves the large sphere (once leaving the large sphere the atom couldn't return back). Results of the simulations permitted evaluation of the following effects:

a) how polarization forces increase the total flux of depositing atoms to the NP;

b) how flux density of the depositing atoms is distributed along the surface of the NP.

## 6. Results of the simulations

Specific calculations were performed to check independence of the results on the large sphere radius R. Also, in the case of a spherical NP, our calculations were compared to predictions of the OML probe theory [9]. Fig.6 depicts calculations for the case A=B (spherical NP) when OML theory can be used. According to this theory, increase of particle flux by the polarization forces is $G/G_0 = 1+V_{float}/V_{oo}$, where $V_{oo}$ is the ion energy far from the particle. The case of β=1 corresponds to $V_{float}/V_{oo}$ equal to two, see definition of the parameter β. Therefore, the $G/G_0$ ratio should be equal to three. Note that this result is the same for any type of attracting potential. We performed calculations with polarization potential ~ $1/r^4$ and with ~ $1/r$ potential[2]. One can see from Fig. 6 that the condition $G/G_0 \approx 3$ is satisfied even with a relatively small R/B ratio ~ 2 for the polarized type of potential. For non-spherical NP, stabilization of the $G/G_0$ ratio occurs starting from R/B ~ 5, see Fig. 7. As one might expect, for 1/r potential, stabilization occurs at a larger R/B ratio due to slowly decreasing type of this potential.

[1] One might think of this radius as an atom mean free path.

Most of the computations were performed for R=10B. While a larger sphere doesn't increase accuracy it tremendously increases the computational time: the NP is seen from a large sphere at a small solid angle, hence, the vast majority of the computed trajectories miss the NP.

Fig. 8 displays the effect of polarization force for NPs of different shapes, from spherical to elliptic with 3:1 semi-axis ratio. This picture answers the question posed in the beginning of this note: whether a long NP has higher or lower deposition rate compared to a spherical one. The answer is - lower. As also seen from this plot, for a small NP the effect of polarizing force can be quite substantial, especially at the early stages of their growth.

The above results show that the effect of polarizing forces becomes less pronounced as a nanoparticle becomes more elongated. However, the total deposition level represents contributions of both the tips of the NP and its "waist". Since electric field gradients are larger at the tips, one might expect predominate deposition at the tips. The Monte Carlo method allows one not only to obtain the average flux density of the depositing atoms. It makes it possible to find the spatial distribution of the flux along the NP surface.

In order to find the distribution of the deposition rate over the NP surface, each hemisphere of the NP was divided into five zones, as shown in Fig.2. The deposition rate at each zone was monitored, see. Fig.9.. The figure shows increase in deposition rate at each zone due to the act of polarization forces. For spherical NP, obviously, deposition rate was equally increased at each zone. For elongated NP, atoms were depositing predominately at the ends of the NP.

---

[2] With this type of potential, the problem simulates an ion motion toward the charged NP. In this case, the right hand side of (1) becomes $\beta/r^2+\mu^2/r^3$. The ion motion has spherical symmetry. The parameter $\beta$, as before, is the ratio of the NP potential to the thermal energy of the ion far away from the NP.

A few computational runs were made with normal distribution of initial velocities. For that purpose, three normally distributed numbers representing three components of the start velocity ($v_x$, $v_y$, and $v_z$) were generated. The results of calculations are shown in Fig.10. It can be seen from this figure that single velocity approximation and maxwellian distribution of the start velocities give very close results. This result can be explained by examining the graph of the critical impact parameter as a function of the atom start velocity (Fig.5). This graph is close to linear around the thermal speed. Therefore, as much as the flux looses at high velocities, it gains at low velocities.

Finally, it should be noted that the uniform accommodation coefficient of the atoms was assumed: it doesn't depend on the atom speed and is equal to unity, i.e. once an atom hits the NP it is "glued" to its surface. In reality, there could be a chemical reason for different accommodation coefficients at different locations at the NP: Open chemical bonds at the tip of the NP favor atoms accommodating at the NP tip. Surface diffusion of the deposited atoms can act in the same direction.

## Conclusions

The Monte Carlo method was applied to simulate trajectories of atoms (or molecules) deposing on a nanoparticle in low-temperature plasma used to produce nanoparticles. The effect of polarization forces pulling atoms into regions of stronger electric field closer to the charged NP was considered. Trajectories of the atoms/molecules were calculated in a straightforward way without using the orbit motion limited (OML) approximation. Different shapes of NPs were considered: from spherical to elliptic with 3:1 semi-axis ratio. The simulations showed that:

1. the effect of polarization forces is substantial for small NPs;
2. the effect of an atom's polarization on the rate of deposition is lower for elongated NPs than for the spherical ones;
3. while a nanoparticle grows, atoms predominately deposit at the NP's tips.

## Acknowledgements

The authors thank Mikhail Shneider (Princeton University, NJ), Yevgeny Raitses (PPPL, NJ) and Igor D. Kaganovich (PPPL, NJ) for multiple fruitful discussions. Help by Andrei Khodak (PPPL, NJ) is highly appreciated.

The research is funded by the U.S. Department of Energy (DOE), Office of Fusion Energy Sciences.

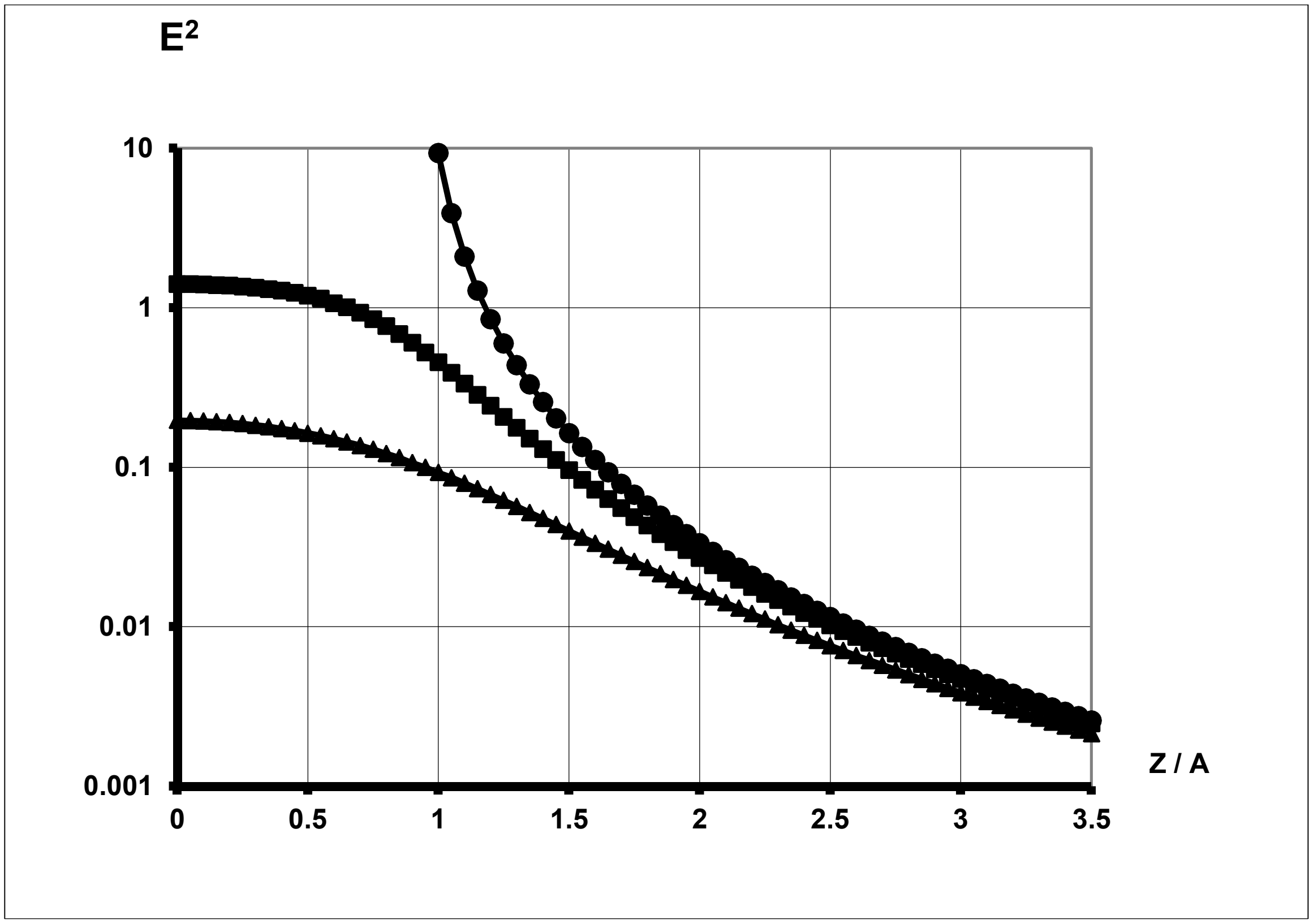

Fig. 1. Axial distribution of the $E^2$ at different distances from the center of the ellipsoid: at the center (circles), at the distance equal to B, the small semi-axis, (squares) and at distance equal to 2B, the small semi-axis doubled (triangles). One can see that the force is directed to the center of the ellipsoid (z=0) and toward the tip of it. Semi-axis ratio is 2:1.

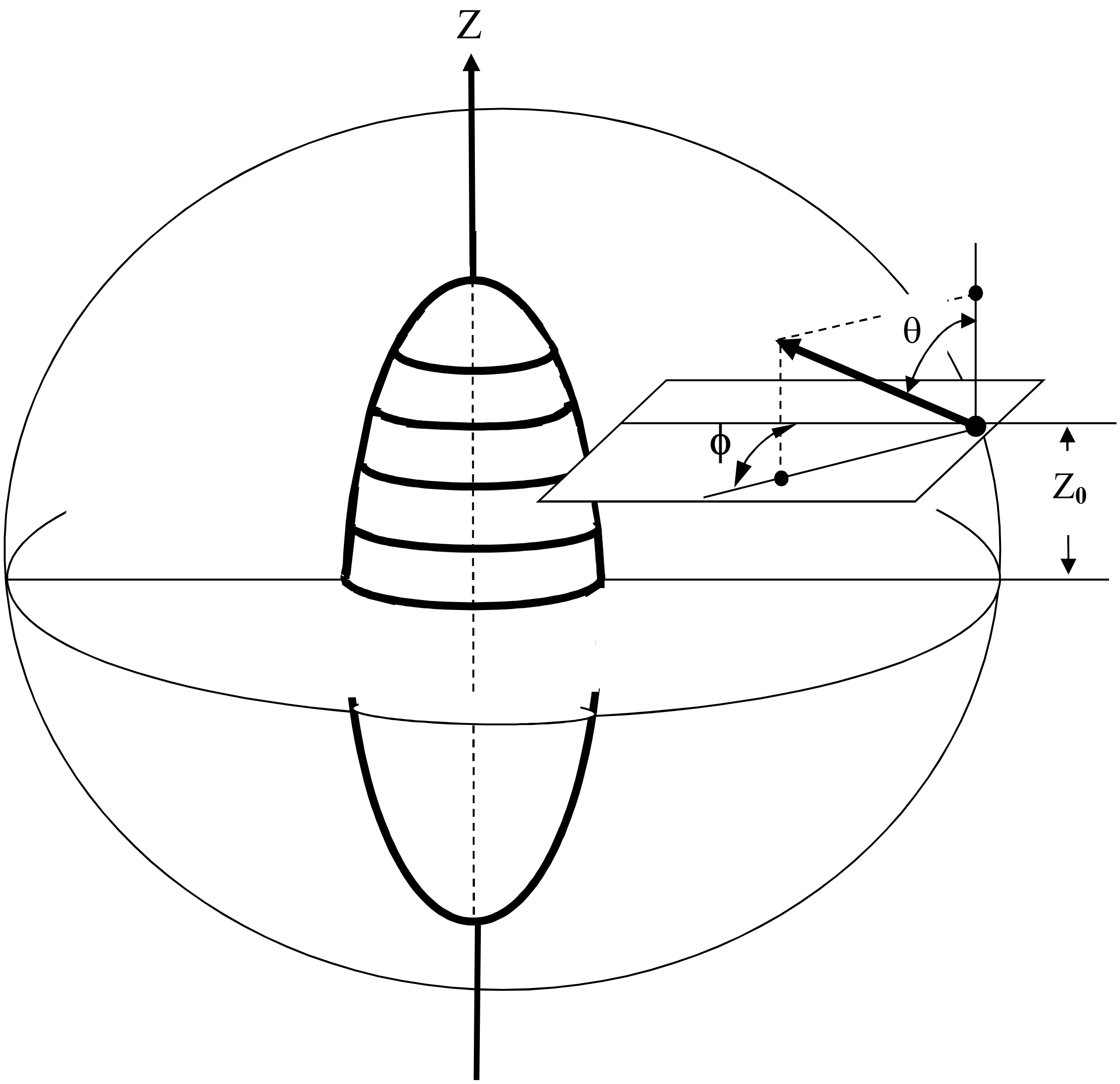

.

Fig. 2. Geometry used in the calculations. Axial coordinate $Z_0$ and the Euler angles that determine the start point and start velocity of the atom entering the computational domain. The five zones of the upper semi-ellipse (NP) used to find distribution of the depositing atoms are shown.

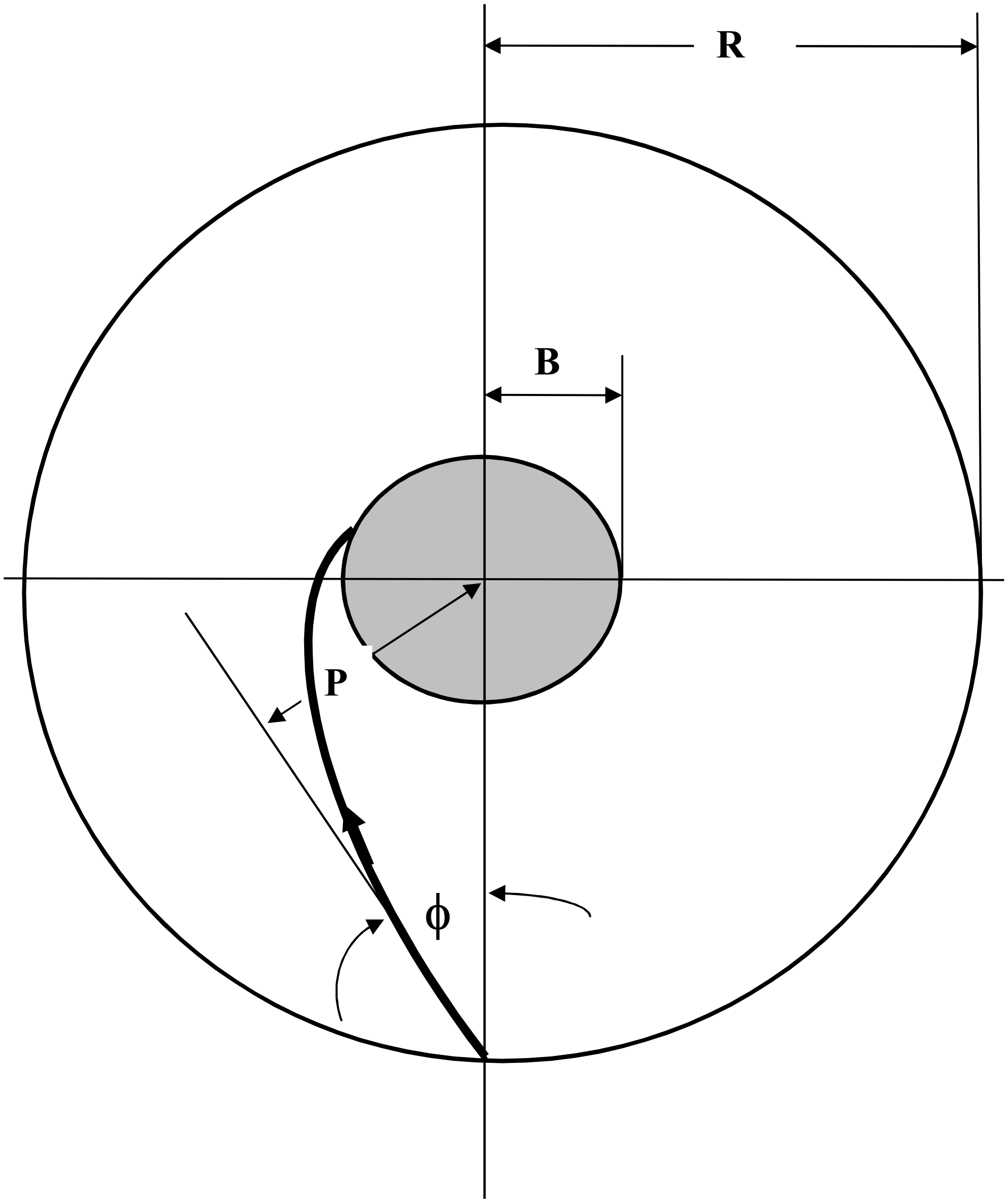

Fig. 3. Atom trajectory in z=0 plane (schematically). Impact parameter P is shown.

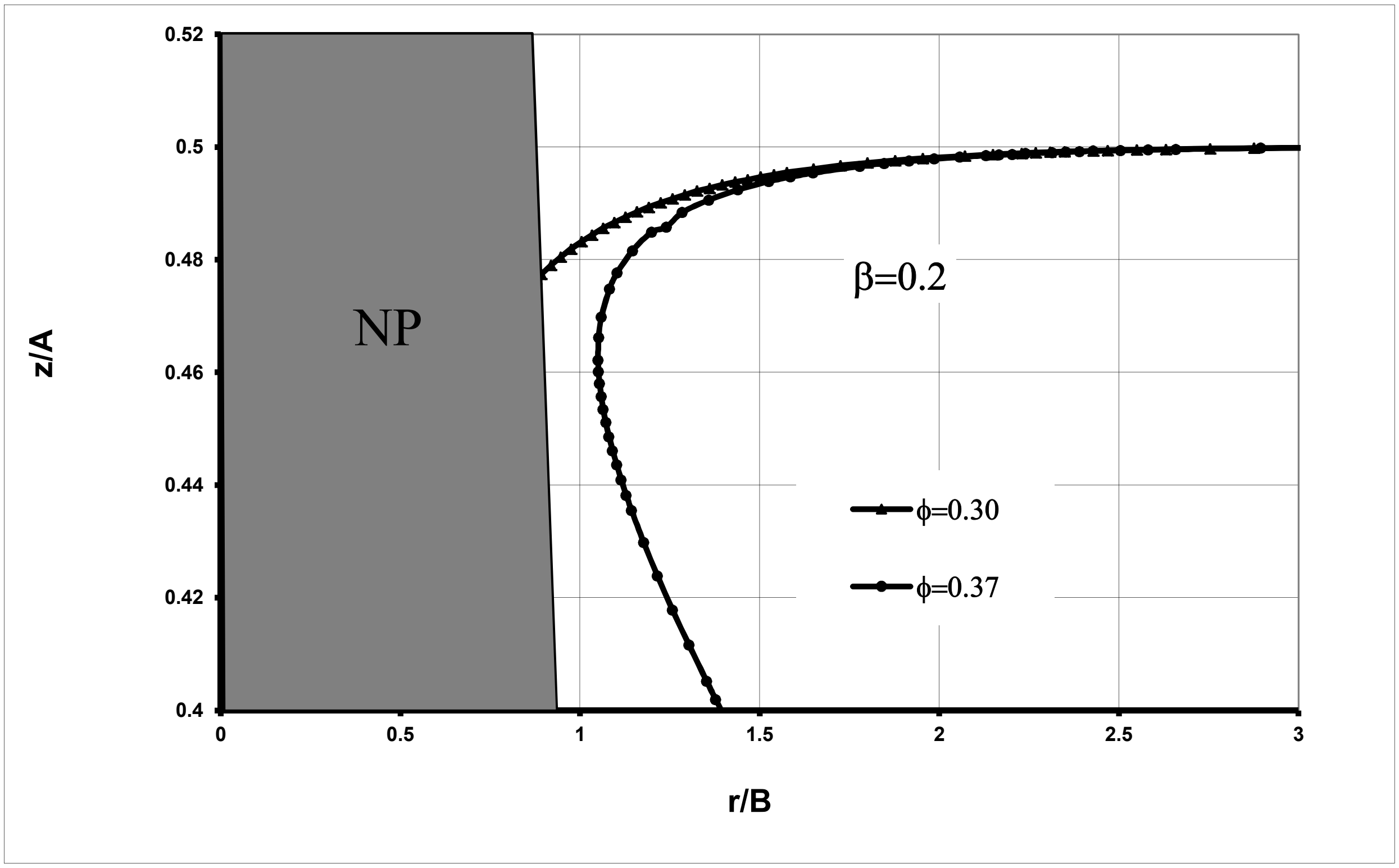

Fig. 4. Two trajectories started at $z_0 = A/2$ with no axial velocities and with different impact parameters. The area occupied by the NP is shown in gray. The atom with low non-dimensional impact parameter ($\phi$=0.3, triangles) hits the NP. The atom with larger impact parameter ($\phi$=0.37, circles) misses it.

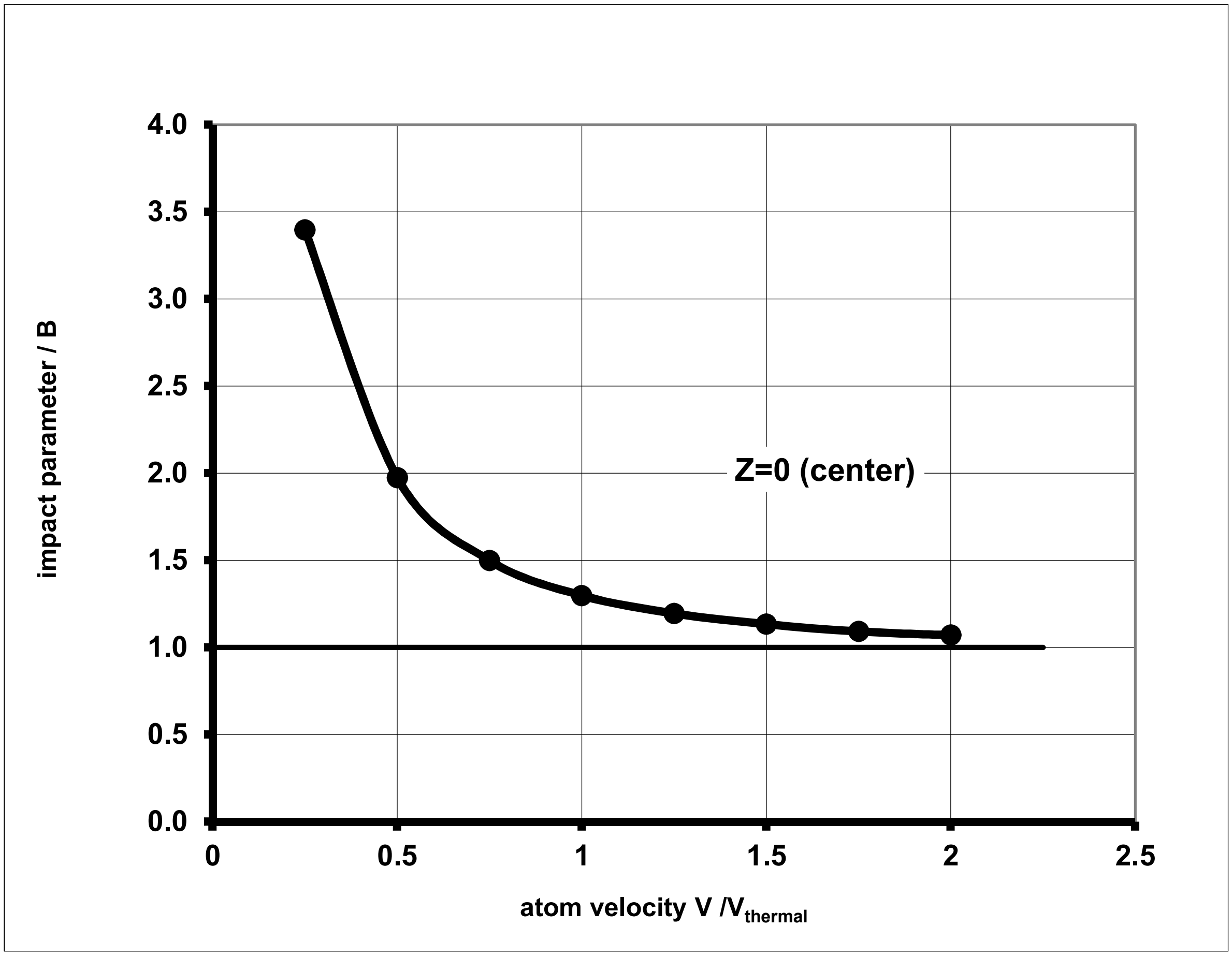

Fig.5. Impact parameter, which separated trajectories that hit the NP from those that do not reach it. Dependence on the atom velocity far from the NP. $\beta = 0.2$. The NP is hit only by those fast atoms which have velocities directed within a small angle $\phi$.

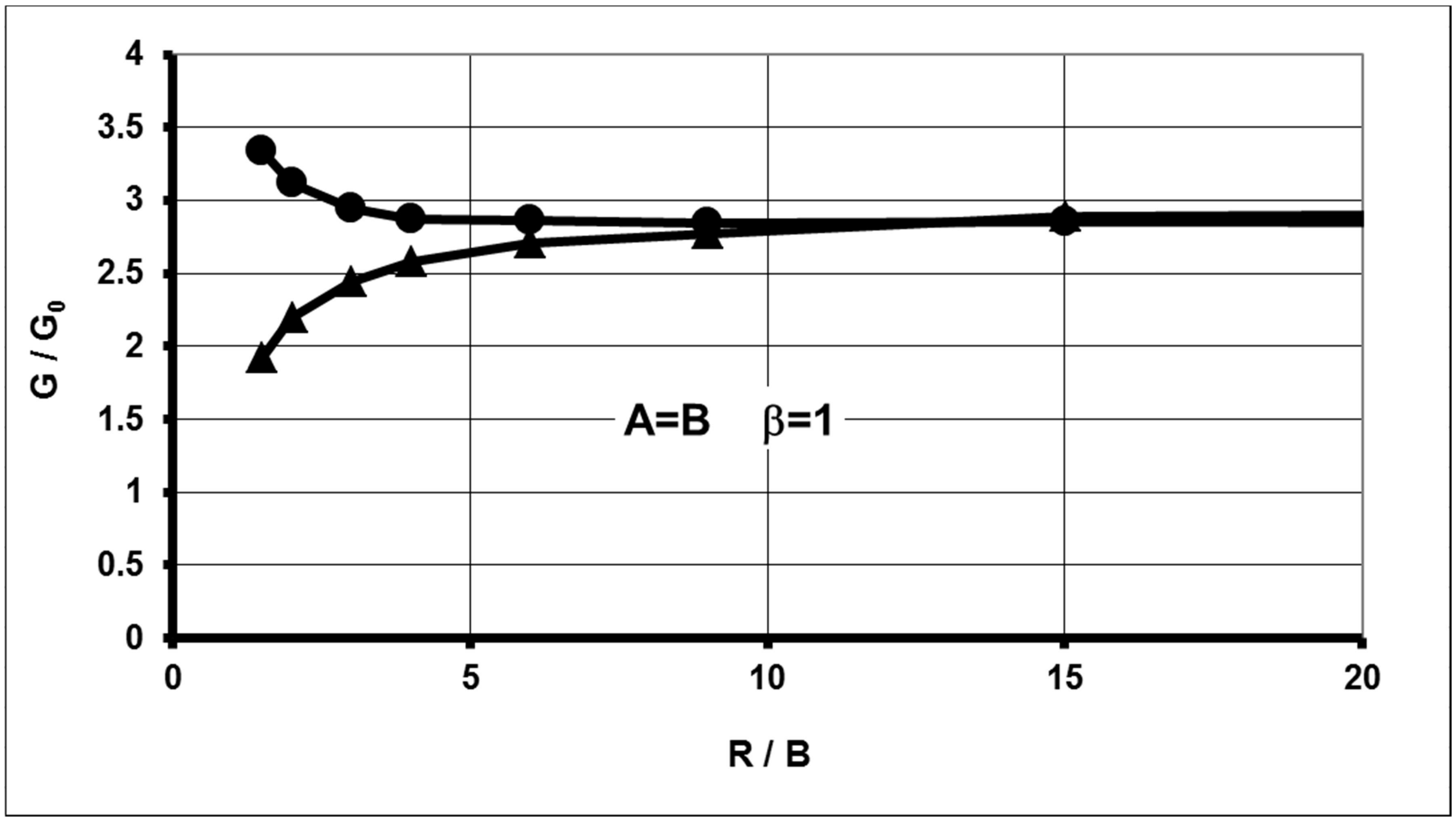

Fig. 6. Increase in the deposition rate as function of the radius of the large sphere R. Circles: polarization potential ~ $1/r^4$. Triangles: Coulomb potential ~ 1/r.

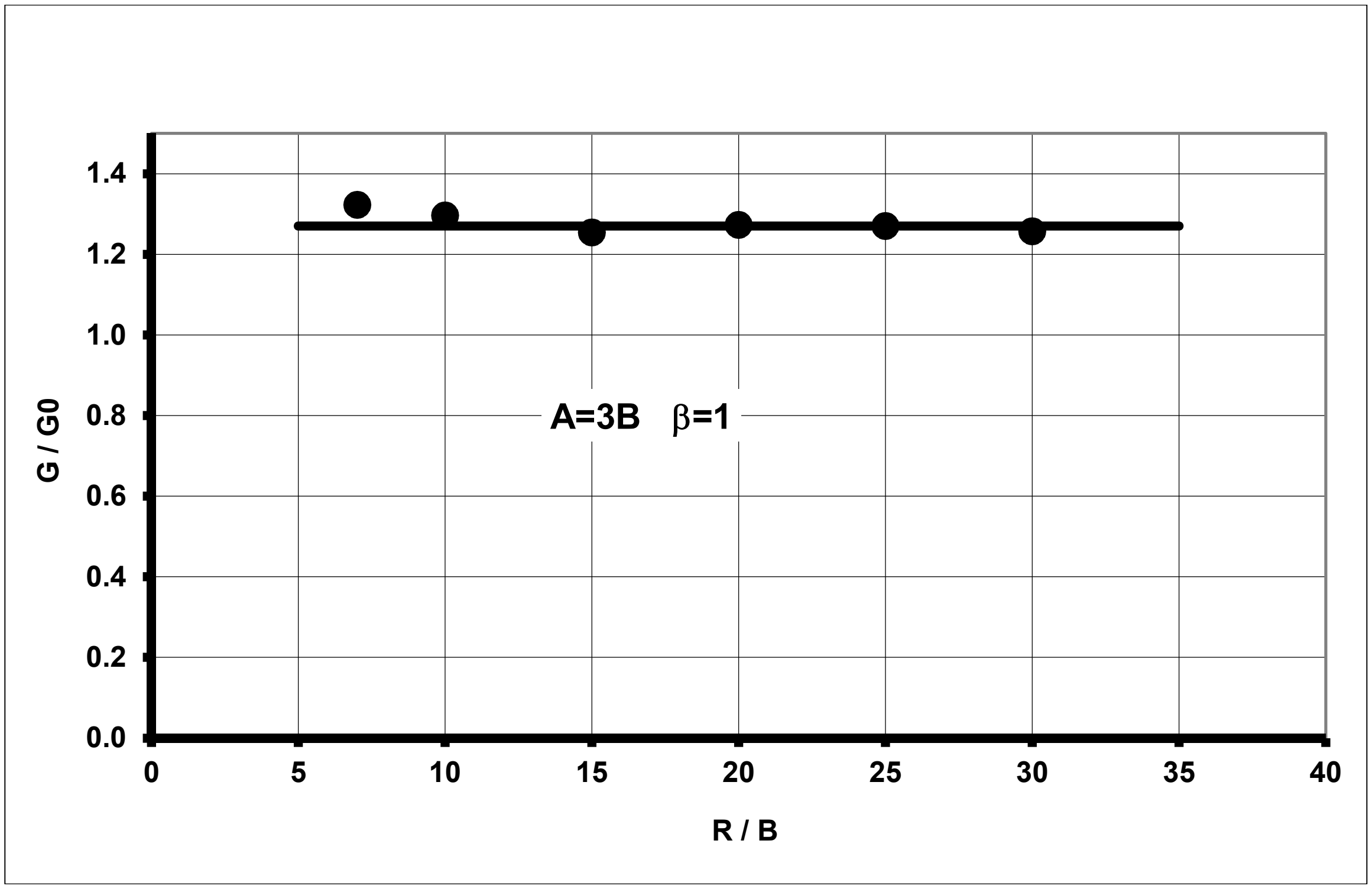

Fig. 7. Dependence of the deposition rate on the radius of the large sphere R: from R/B ratio over approximately 15, the result ceases to depend on R.

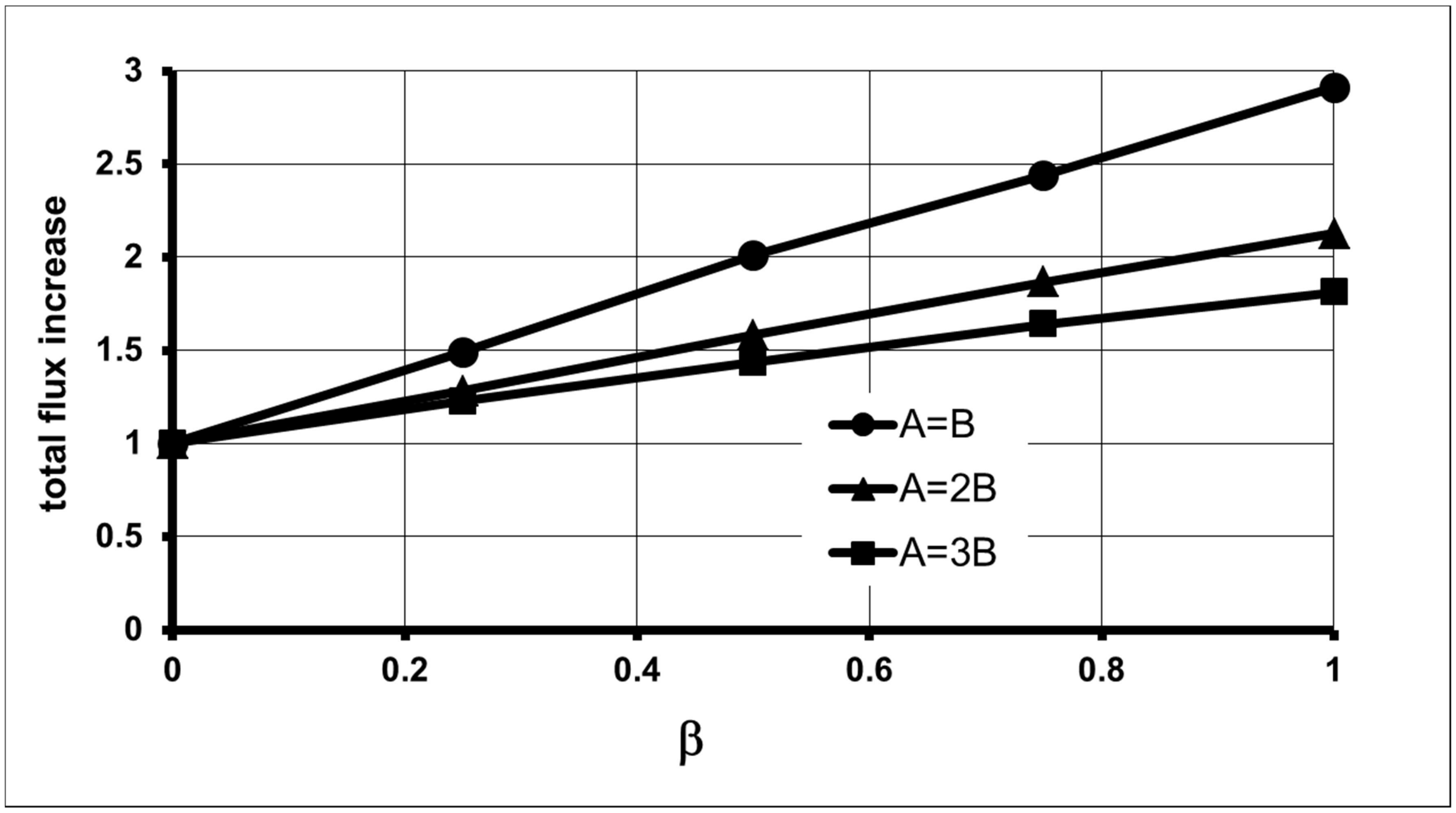

Fig. 8. Ratio of the two deposition rates: with polarizing force and without it for various NP semi-axis ratios: 1:1 (circles), 2:1 (triangles), 3:1 (squares).

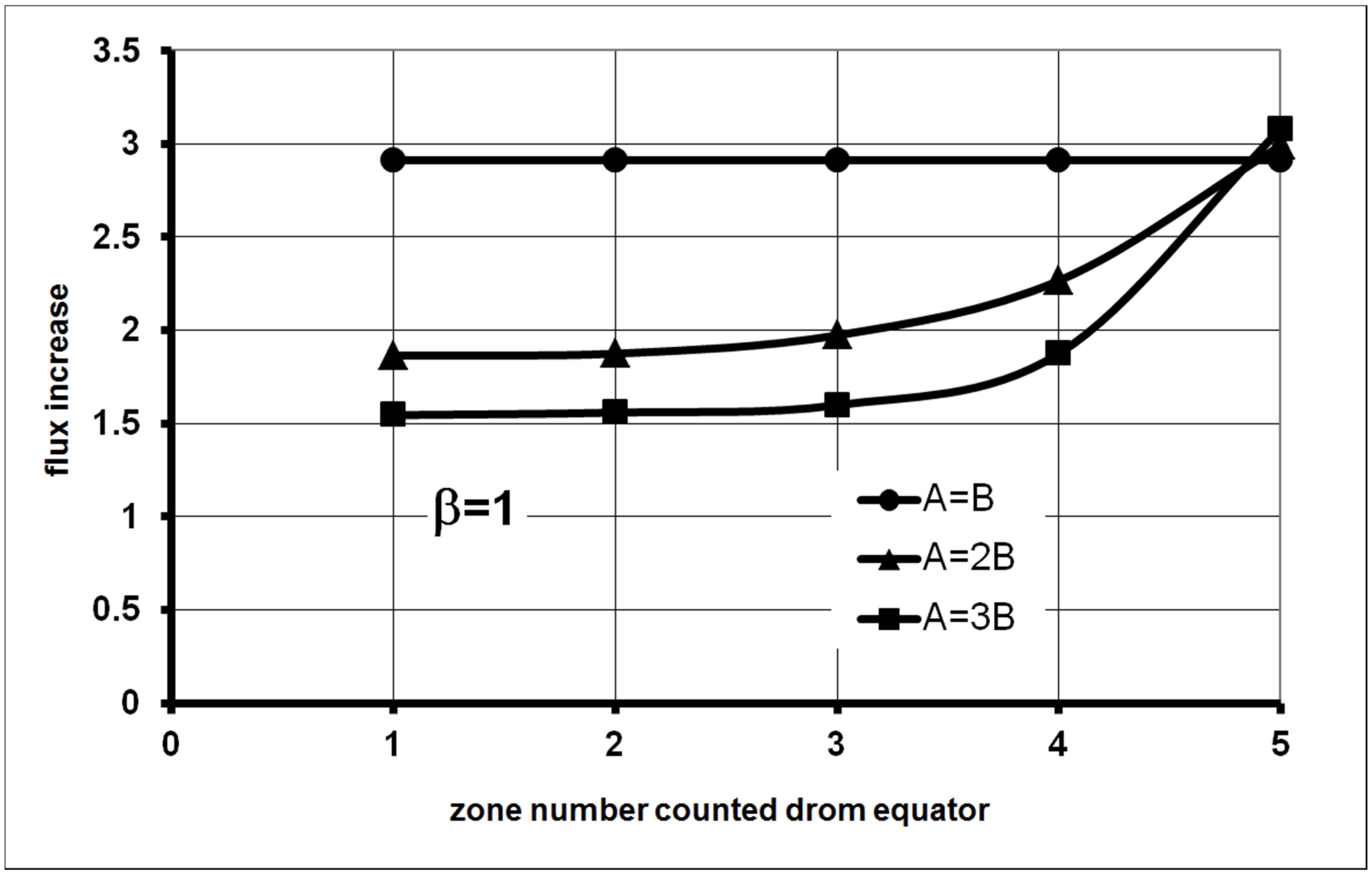

Fig. 9. Spatial distribution of the increase in the deposition rate. It is uniform for spherical NP. It is maximal at the tip of the elongated NP. Decrease of the total deposition rate – see Fig. 8.

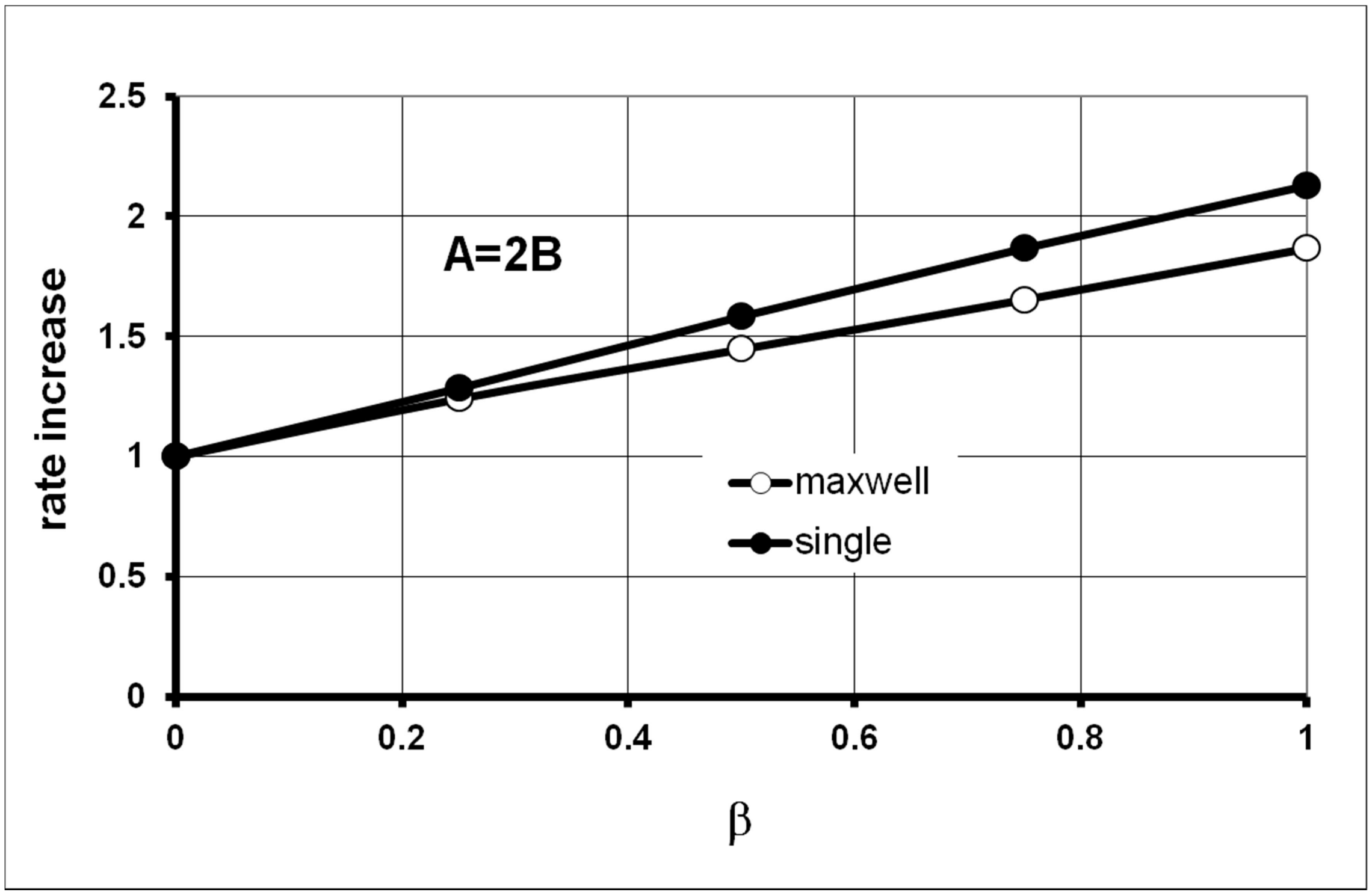

Fig. 10. Comparison of increase in deposition rate calculated with a single start velocity approximation and with maxwellian distribution of the start velocity.